# The optimal vector phase matching conditions in biaxial crystalline materials determined by extreme surfaces method: the case of orthorhombic crystals

Oleh Buryy[1], Dmytro Shulha[2], Nazariy Andrushchak[3], Andriy Danylov[2], Bouchta Sahraoui[4], Anatoliy Andrushchak[2]

[1]Department of Semiconductor Electronics, Lviv Polytechnic National University, Lviv, Ukraine
[2]Department of Applied Physics and Nanomaterials Science, Lviv Polytechnic National University, Lviv, Ukraine
[3]Department of Computer-Aided Design Systems, Lviv Polytechnic National University, Lviv, Ukraine
[4] University of Angers, Institute of Sciences and Molecular Technologies of Angers, Angers, France

**The optimal geometries of vector phase matching are determined for the cases of second harmonic, sum and difference frequency generation in a number of orthorhombic nonlinear optical crystals – KTP, KTA, KB5, $KNbO_3$, LBO, CBO, LRB4. Extreme surface method was used to define wave vectors directions of highest possible generation efficiency. As it is shown, in a significant number of cases vector phase matching ensures higher efficiencies than the scalar one.**

## INTRODUCTION

In our previous paper [1] we determined highest efficiencies of nonlinear effects, i.e., second harmonic generation (SHG), sum frequency generation (SFG), difference frequency generation (DFG) in several uniaxial crystals for the case of vector phase matching (PM). Our approach is based on the construction and analysis of special surfaces called extreme that are representing maximal values of generation efficiency for all directions of the output light beam determined after optimization on the directions of pump (initial) beams (see for example our works [8-13], where extreme surfaces method has been used for electro-, piezo- and acousto-optical effects in crystalline materials). Here we use it for solving the same problem for more complex cases of biaxial crystals. Although the problem of scalar PM in biaxial crystals has already been studied in a number of papers [8-12], to our knowledge, the case of vector PM has not been explored to the same extent yet. We decided to limit scope of analysis to crystals of orthorhombic and monoclinic (in the continuation of this paper) syngonies; although SHG was revealed or expected to be revealed in some triclinic organic crystals [13–16], there is no information about their nonlinear susceptibilities. Nevertheless, some preliminary results for SHG was published in [17,18], here they are also provided for greater completeness and comparison. For all crystals, results obtained for vector PM are compared with the ones obtained for scalar PM when the directions of pump and output beams coincide.

## BASIC RELATIONS

The main relations used in our consideration are the same as the ones of [1,17,18]. Particularly, to consider only the geometrical factor of interactions the efficiency is determined by the expression

$$\eta = \frac{\left(\vec{e}_3 \hat{d} \vec{e}_1 \vec{e}_2\right)^2}{n(\lambda_1) n(\lambda_2) n(\lambda_3)}, \quad (1)$$

where $\lambda_i$ are the wavelengths, $\hat{d}$ is the tensor of nonlinear coefficients, $n(\lambda_1)$, $n(\lambda_2)$, $n(\lambda_3)$ are the refraction indices of pump (1, 2) and output (3) beams, $\vec{e}_1$, $\vec{e}_2$, $\vec{e}_3$ are the unit vectors parallel to the electric vectors of the corresponding waves, $\vec{e}_j = \hat{\varepsilon}\vec{i}_j$, $\hat{\varepsilon}$ is the dielectric permittivity tensor, $\vec{i}_j$ is the electric displacement unit vector (or light wave polarization) determined for each direction of the wave vector $\vec{k}$ in a known manner [19].The expression (1) is used here to construct extreme surfaces representing the highest achievable values of η for all possible directions of the output wave vector $\vec{k}_3$, determined by the angles θ, φ of the spherical coordinate system.

It also should be considered that the highest efficiency of nonlinear interactions is achieved if PM condition is realized. In general case of vector PM it is:

$$\vec{k}_3 = \vec{k}_1 \pm \vec{k}_2, \qquad (2)$$

where the upper sign corresponds to SHG or SFG and the lower – to DFG. For scalar PM this condition transforms to the one between the absolute values of the wave vectors, $k_3 = k_1 \pm k_2$.

Except for the directions of optic axes, for each direction in non-cubic crystal two light waves with orthogonal polarizations can propagate. For biaxial crystals the lengths of their wave vectors $k$ are determined from the equation [14]:

$$k^4\left(K_1^2 m_1^2 + K_2^2 m_2^2 + K_3^2 m_3^2\right) - k^2\left(K_1^2\left(K_2^2 + K_3^2\right)m_1^2 + K_2^2\left(K_1^2 + K_3^2\right)m_2^2 + K_3^2\left(K_1^2 + K_2^2\right)m_3^2\right) + K_1^2 K_2^2 K_3^2 = 0, \qquad (3)$$

where $K_1$, $K_2$, $K_3$ are the lengths of wave vectors along crystallographic axes, $K_i = 2\pi N_i/\lambda$, $N_i$ are the main refraction indices, $i$ = 1,2,3, $m_1$, $m_2$, $m_3$ are the components of the wave normal. The equation (3) describes the double-cavity wave vector surface (Fig. 1). The external part of this surface corresponds to a 'slow' ($s$) wave characterized by higher value of the refraction index $n$ and, consequently, by lower value of light velocity and higher absolute value of the wave vector, $k = 2\pi n/\lambda$. Contrary, the internal part of this surface corresponds to a 'fast' ($f$) wave characterized by lower values of refraction index and wave vector length. The PM condition (2) can be satisfied for SHG and SFG in biaxial crystals in two possible cases: (i) both pump waves are slow (*ssf* or type I phase matching) or (ii) one pump wave is slow and the other is fast (*sff* or type II phase matching) [20]; the output wave is fast in both cases. For DFG in biaxial crystal type I phase matching is *ffs* and the type II one is *fss* or *fsf*.

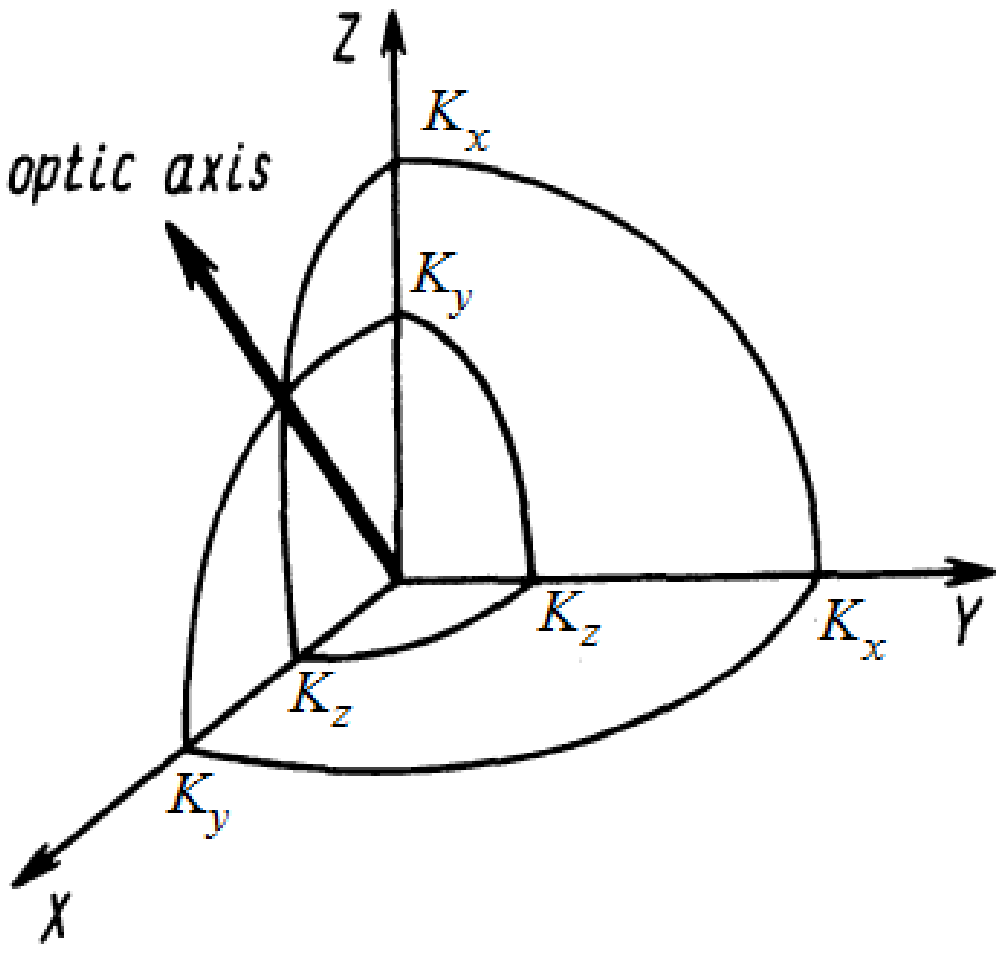

Fig. 1. One-eighth part of the wave vector surface for the case of biaxial crystal.

For definiteness, consider the case of *ssf* phase matching. As it is seen from (2), vector PM condition for each $\vec{k}_3$ can be satisfied for a set of $\vec{k}_1$, $\vec{k}_2$ vectors ending on line $C$ in Fig. 2. The position of this line can be determined in the following manner. Let's construct a wave vector surface (only its external 'slow' part) for $\vec{k}_1$ from the origin and the one for $\vec{k}_2$ from the end of the vector $\vec{k}_3$. If PM is possible, these surfaces must intersect along some closed line. This line belongs to both wave vector surfaces, so the equality $\vec{k}_1 = \vec{k}_3 + \vec{k}_2$ should be satisfied in each point of it. Because of the symmetry of wave vector surfaces in relation to inversion, this equality is equivalent to $\vec{k}_3 = \vec{k}_1 + \vec{k}_2$. Thus, this line corresponds to line $C$ in Fig. 2. Generally, for the same $\vec{k}_3$, different values of the efficiency η correspond to different $\vec{k}_1$, $\vec{k}_2$. To determine the maximal efficiency for given $\vec{k}_3$, $\eta_{max}(\vec{k}_3)$, we calculated the values of η for all $\vec{k}_1$, $\vec{k}_2$ (certainly, with some small enough step) and compared the obtained results. No additional optimization technique was used for searching of $\eta_{max}(\vec{k}_3)$. As well as in [1], determination of points of line *C* is based on Dragilev' method, except that for the case of uniaxial crystals this line can be determined as the line of intersection of spheres or ellipsoids, whereas for biaxial ones it is the line of intersection of different parts of wave vector surfaces (internal of external depending on the type of PM). Simultaneously with the sequential determination of all points of the line *C* the efficiency η is calculated and, based on obtained results, the maximal one, $\eta_{max}(\vec{k}_3)$ is determined.

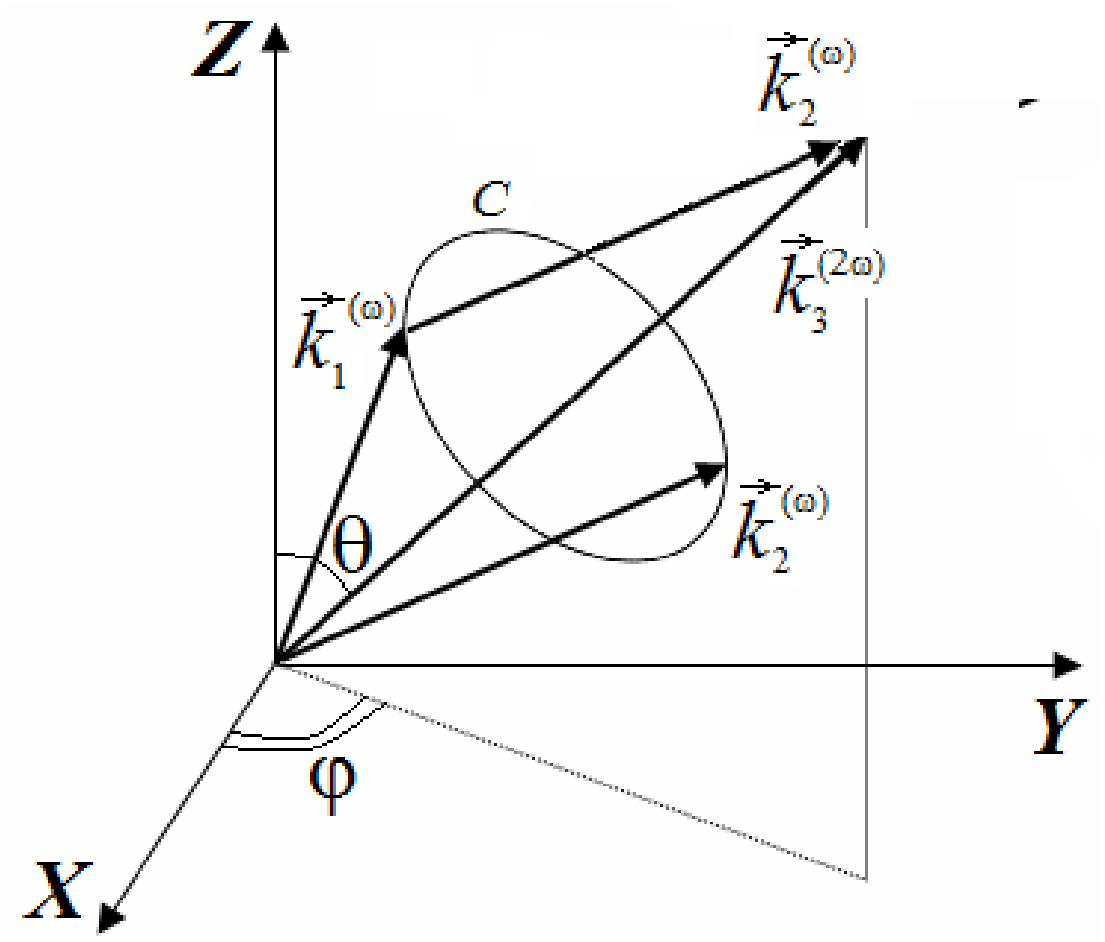

Fig. 2. The mutual position of the wave vectors of pump and output beams.

After the maximal efficiency $\eta_{max}(\vec{k}_3)$ was determined for all possible wave vector $\vec{k}_3$ directions (θ = 0…π, φ = 0…2π), the global maximal value of the efficiency $\eta_{max}^{extr}$ can be found as the highest value from the set of the values of $\eta_{max}$ determined for all $\vec{k}_3$. The dependence $\eta_{max}(\theta,\varphi)$ can be conveniently presented as a 3D surface, the extreme one in accordance with the method of its calculation, i.e. searching the maximum for each its point. Here such surfaces are constructed, and optimal PM conditions are determined for orthorhombic nonlinear optical crystals $KTiOPO_4$ (KTP), $KTiOAsO_4$ (KTA), $KB_5O_8 \cdot 4H_2O$ (KB5), $KNbO_3$, $LiB_3O_6$ (LBO), $CsB_3O_5$ (CBO), $LiRbB_4O_7$ (LRB4), parameters of which are given in Table 1. Values of the parameters were taken from [20, 21] the refraction indices were calculated in accordance with Sellmeier equations given in [21]. The transformation of the axes from the crystal-physics coordinate system ($X_cY_cZ_c$) to the crystal-optics one (*XYZ*) were carried out in accordance with the rules given in [21, 22]. Note that all extreme surfaces were constructed in crystal-optics coordinate system.

**Table 1.** Parameters of the considered crystals

| **Crystal** | Correspondence between the crystal-physics and crystal-optics coordinate systems | **Nonlinear susceptibilities, $d_{ij}$, pm/V** |
|---|---|---|
| **Point group mm2** | | |
| KTP | $X_cY_cZ_c \leftrightarrow XYZ$ | $d_{15}$ = 1.9; $d_{24}$ = 3.7; $d_{31}$ = 2.2; $d_{32}$ = 3.7; $d_{33}$ = 14.6 |
| KTA | $X_cY_cZ_c \leftrightarrow XYZ$ | $d_{15}$ = 2.5; $d_{24}$ = 4.4; $d_{31}$ = 2.9; $d_{32}$ = 5.1; $d_{33}$ = 16.2 |
| KB5 | $X_cY_cZ_c \leftrightarrow XYZ$ | $d_{15}$ = $d_{31}$ = 0.04; $d_{24}$ = $d_{32}$ = 0.003; $d_{33}$ = 0.05 |
| $KNbO_3$ | $X_cY_cZ_c \leftrightarrow YXZ$ | $d_{15}$ = -12.4; $d_{24}$ = -12.8; $d_{31}$ = -11.9; $d_{32}$ = -13.7; $d_{33}$ = -20.6 |
| LBO | $X_cY_cZ_c \leftrightarrow XZY$ | $d_{15}$ = $d_{31}$ = -0.67; $d_{24}$ = $d_{32}$ = 0.85; $d_{33}$ = 0.04 |
| **Point group 222** | | |
| CBO | $XYZ \leftrightarrow ZXY$ | $d_{14}$ = $d_{36}$ = $d_{25}$ = 1.49 |
| LRB4 | $X_cY_cZ_c \leftrightarrow YZX$ | $d_{14}$ = $d_{36}$ = $d_{25}$ = 0.45 |

## RESULTS AND DISCUSSION

*Second-harmonic generation*

The wavelengths of pump beams used in the calculations of the maximal achievable SHG efficiency are equal to 1.0642 μm for all crystals except for KB5 for which it is equal to 0.5321 μm, because at this wavelength the nonlinear susceptibilities were determined [21].

The general and the top views of extreme surfaces of the SHG efficiency $\eta_{max}$ for the investigated crystals are shown in Figs. 3, 4. The black lines on the Figures correspond to the scalar PM conditions. It should be noted that the point (0;0;0), as well as the lines connecting this point with the edges of extreme surface, do not belong to this surface and appear in the figures only in connection with the method of 3D surfaces constructing in the software used.

Because type II PM can not be achieved in KTA, KB5 and $KNbO_3$ crystals at considered wavelengths, only one extreme surface (for *ssf* PM) is shown for each of them. The results of the optimization are given in Table 2. For brevity, the angle values listed in the Table correspond to only one of the equivalent maxima. The values shown in the parentheses in the last column of Table 2 are the relative increase of SHG efficiency caused by vector PM using in relation to the scalar one, $\gamma = \left(\eta_{vect}^{extr} - \eta_{scal}^{extr}\right) / \eta_{scal}^{extr}$ (in percents).

As it is seen from Figs. 3, 4, the forms of extreme surfaces for the crystals of the same point group are usually not similar (such similarity is observed only for the cases of *ssf* PM in KTA and KTP) that is obviously caused by different rules of coordinates transformation and different relationships between the values of nonlinear susceptibilities $d_{ij}$. Contrary, in the case of KTP and KTA the coordinates transformations are the same and the relationships between $d_{ij}$ are, in general, similar. Particularly, the coefficients $d_{33}$ for KTP and KTA are several times higher than the other ones. Although the transformation rule for KB5 crystal is the same as the ones for KTP and KTA, all nonlinear susceptibilities $d_{ij}$ are commensurate for it, so it is not surprising that the extreme surface for KB5 differs from the ones for KTP and KTA. It should be emphasized that the extreme surfaces for *ssf* PM in KTA and KTP have not got the symmetry axes of $4^{th}$ order as it can be mistakenly concluded from Fig. 3. In particular, for KTA crystal, the angle in XY plane between the directions corresponding to the maxima of $\eta_{max}$ is equal to 93.5° between the 'petals' placed in the first and the second as well as in the third and the fourth octants. Correspondingly, the angles between the 'petals' placed in the second and the third, the fourth and the first octants is 86.5°. The same angles for KTP crystal are 96.5° and 83.5°, so the extreme surfaces do not reveal symmetry higher than orthorhombic.

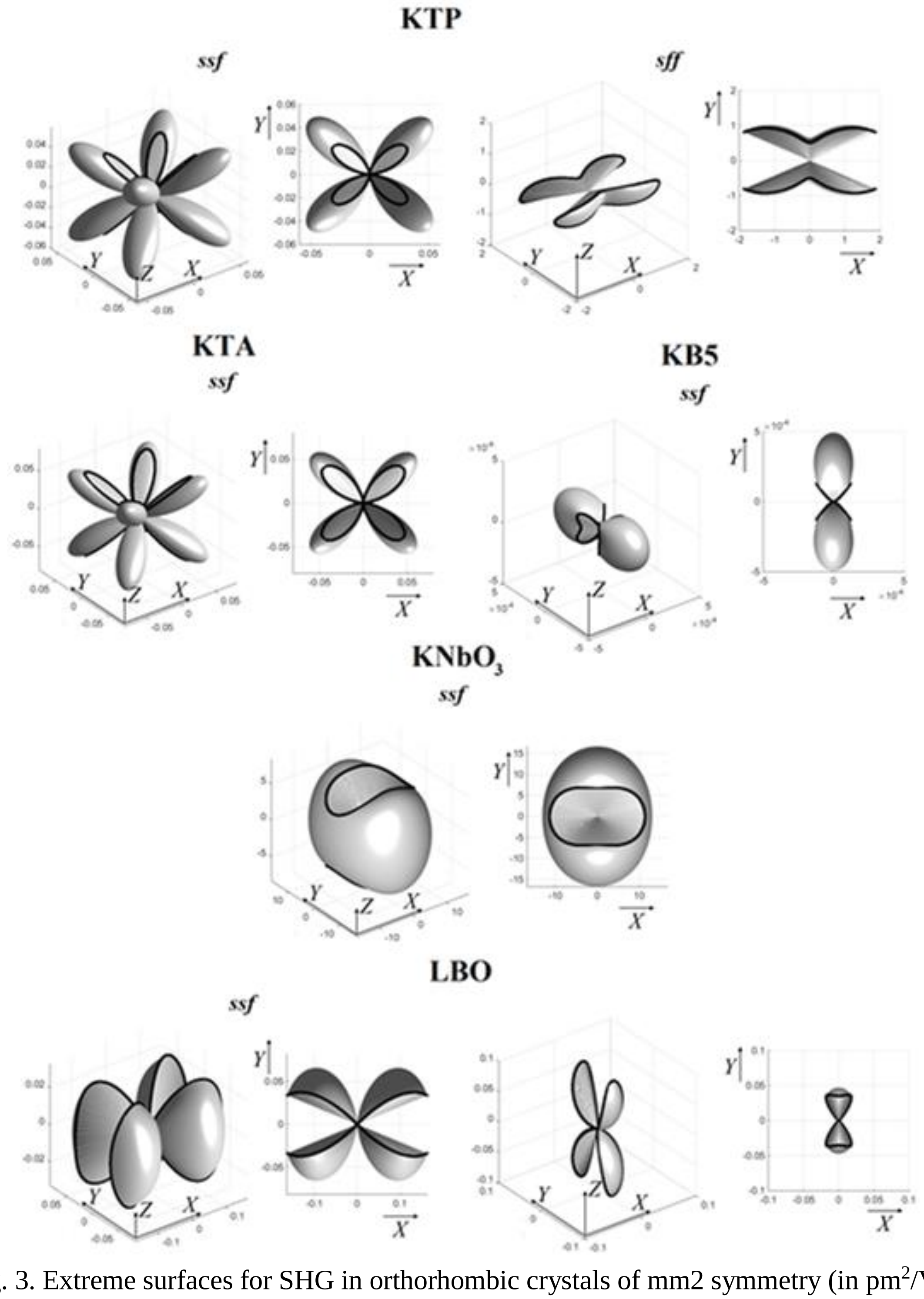

Fig. 3. Extreme surfaces for SHG in orthorhombic crystals of mm2 symmetry (in pm$^2$/V$^2$).

For some crystals (KTA, KB5, $KNbO_3$), *sff* PM conditions are not fulfilled for the considered wavelengths, so these crystals are not mentioned in the appropriate part of Table 2. Also, as it is followed from our calculations, the use of type I (*ssf*) vector PM practically does not allow to increase the efficiency $\eta$ in comparison with the case of scalar PM for LBO, CBO and LRB4 crystals. In these cases, the lines corresponding to scalar PM in figures frame the edges of extreme surfaces and pass through the points corresponding to $\eta_v^{extr}$. The same situation takes place in the case of type II (*ssf*) vector PM in KTP, LBO and CBO. In other cases, the SHG efficiency ensured by *ssf* vector PM is essentially higher than the one for scalar PM for KB5 (145%), KTP (49%), $KNbO_3$ (39%) and KTA (17%), whereas for *sff* vector PM it is remarkably lower: for LRB4 it is equal only to 7.4%. Thus in some cases vector PM usage allows to obtain significantly higher efficiencies than by using scalar one, which indicates a possible way to enhance the performance of nonlinear optical devices.

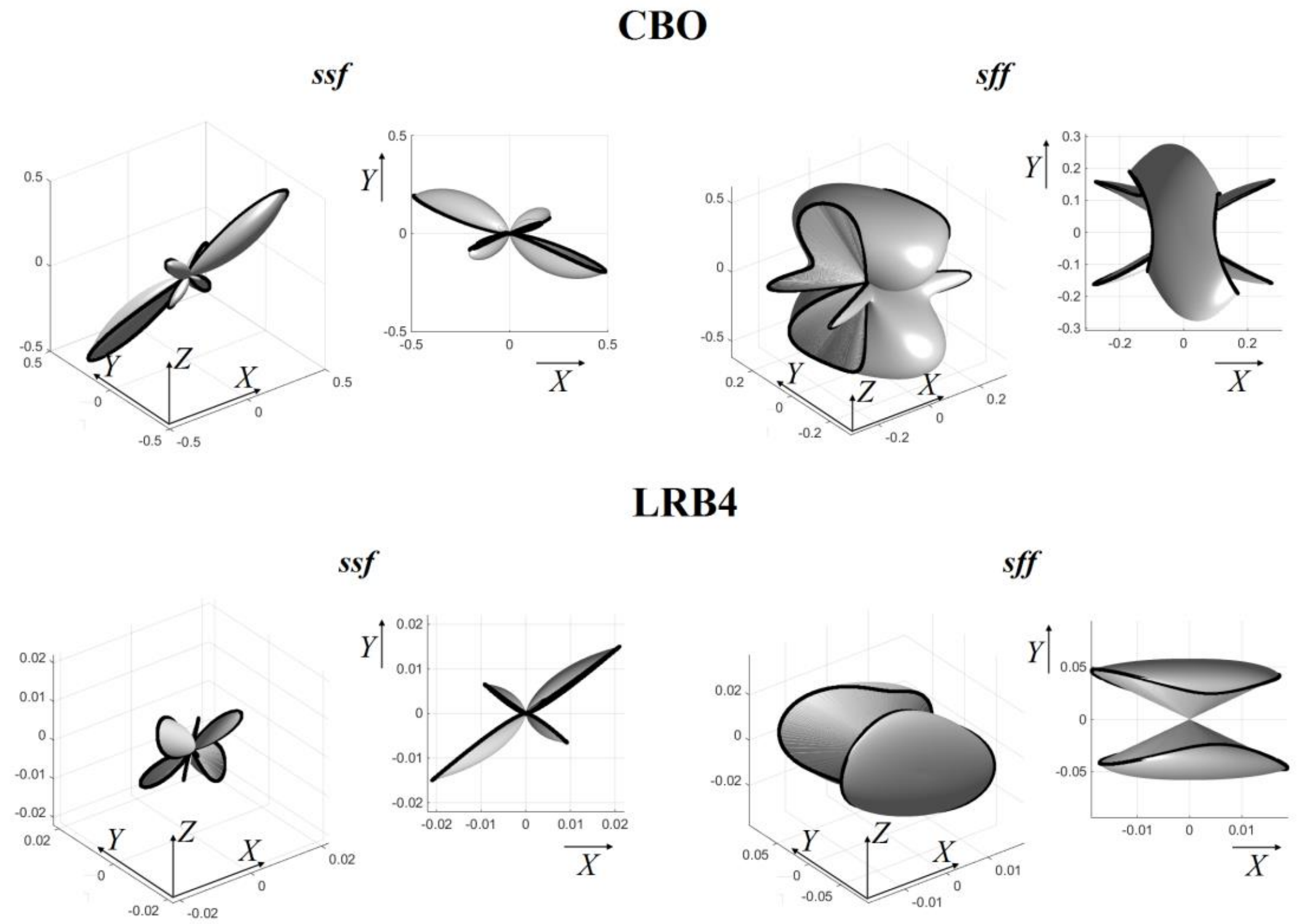

Fig. 4. Extreme surfaces for SHG in orthorhombic crystals of 222 symmetry (in $pm^2/V^2$).

**Table 2.** The maximal achievable SHG efficiencies η and corresponding angular parameters

| Crystal | Scalar PM | | | Vector PM | | | | |
|---|---|---|---|---|---|---|---|---|
| | Angles, deg. | | $\eta_{scal}^{extr}$, $pm^2/V^2$ | Angles (pump beams), deg. | | Angles (SH beam), deg. | | $\eta_{vect}^{extr}$, $pm^2/V^2$ |
| | θ | φ | | $\theta_p$ | $\varphi_p$ | θ | φ | |
| **type I phase matching** | | | | | | | | |
| KTP | 134.7 | 218.4 | 0.055 | 116.0<br>129.3 | 36.3<br>48.1 | 122.7 | 41.7 | 0.082<br>(49%) |
| KTA | 131.6 | 42.1 | 0.081 | 118.7<br>128.2 | 141.2<br>131.7 | 123.5 | 136.8 | 0.095<br>(17.3%) |
| KB5 | 121.4 | 133.4 | $2.0 \cdot 10^{-4}$ | 100.6<br>79.4 | 90<br>90 | 90 | 90 | $4.9 \cdot 10^{-4}$<br>(145%) |
| $KNbO_3$ | 71.4 | 0 | 12.0 | 102.6<br>77.4 | 90<br>90 | 90 | 90 | 16.7<br>(39.2%) |
| LBO | 89.9 | 168.4 | 0.17 | coincides with scalar PM | | | | |
| CBO | 51.4 | 337.9 | 0.67 | coincides with scalar PM | | | | |
| LRB4 | 53.3 | 215.4 | 0.032 | coincides with scalar PM | | | | |
| **type II phase matching** | | | | | | | | |
| KTP | 90 | 156.6 | 2.0 | coincides with scalar PM | | | | |
| LBO | 20.5 | 90 | 0.10 | coincides with scalar PM | | | | |
| CBO | 13.1 | 325.5 | 0.60 | almost coincides with scalar PM* | | | | |
| LRB4 | 69.1 | 109.7 | 0.054 | 89.3<br>83.1 | 91.4<br>90.0 | 86.2 | 90.7 | 0.058<br>(7.4%) |

* *The relative increase of SFG efficiency is about 3% or lower.

*Sum frequency generation*

The results of the efficiency η calculations for some cases of SFG mentioned in [20, 21] are given in Table 3. Because *sff* PM is not achieved in KB5 crystal for any considered wavelengths, only the data for *ssf* PM are indicated for this crystal. The most salient examples of extreme surfaces are shown in Figs. 5,6. As it is seen from the figures, the forms of extreme surfaces are quite diverse. Only in some cases the extreme surfaces for SFG and SHG look similar, namely, for *ssf* PM in KTP, KTA, $KNbO_3$ (the second surface for this crystal shown in Fig. 5) as well as *sff* PM in LBO. Such a similarity should take place where the wavelengths of the beams involved in the SFG process are close to those considered for the SHG.

Along with that, even for the same crystal and the same type of PM, extreme surfaces can be essentally different. By changing the wavelengths of the beams so that the PM condition (2) is fullfilled, we can trace the change in the shape of the extreme surface. For example, in Fig. 7 it is shown the sequence of extreme surfaces that occur for *sff* PM in KTA crystal obtained during a transition from (0.6594; 1.3188; 0.4396) to (1.0642; 1.9079; 0.6831); here the numbers in parentheses indicate the wavelengths of the input (correspondingly, *s* and *f*) and output beams. As can be seen from Fig. 7, at relatively small wavelengths the extreme surface consists of two separate parts, which closes as the wavelength increases; thus, the increase in the wavelength leads to a change of surface topology. Such changes can be also traced for the other crystals considered here.

**Table 3**. The maximal achievable SFG efficiencies η and corresponding angular parameters

| **Crystal** | **$\lambda_1$, μm** | **$\lambda_2$, μm** | **$\lambda_3$, μm** | **Scalar PM** | | | **Vector PM** | | | | |
|---|---|---|---|---|---|---|---|---|---|---|---|
| | | | | **Angles, deg.** | | **$\eta^{extr}_{scal}$, pm²/V²** | **Angles (pump beams), deg.** | | **Angles (SFG beam), deg.** | | **$\eta^{extr}_{vect}$, pm²/V²** |
| | | | | **θ** | **φ** | | **$\theta_p$** | **$\varphi_p$** | **θ** | **φ** | |
| **type I phase matching (*ssf*)** | | | | | | | | | | | |
| KTP | 1.0642 | 1.5918 | 0. 6378 | 37.6 | 214 | 0.044 | 63.6<br>47.5 | 33<br>50 | 57 | 39 | 0.14<br>(218%) |
| | 0.76 | 1.77 | 0.5321 | 40.9 | 215.7 | 0.044 | 65.4<br>48.7 | 146.9<br>127.6 | 60.3 | 142 | 0.17<br>(286%) |
| | 1.90768 | 2.40688 | 1.0642 | 32.8 | 31.1 | 0.033 | 62.7<br>32.5 | 45.8<br>50 | 55 | 39.2 | 0.14<br>(324%) |
| KTA | 0.6594 | 1.3188 | 0.4396 | 121 | 44.2 | 0.054 | 67.9<br>60.1 | 138.3<br>128.6 | 65.2 | 135.3 | 0.075<br>(38.9%) |
| | 1.0642 | 1.9079 | 0.6831 | 36.7 | 143.9 | 0.059 | 62.7<br>33.2 | 47.4<br>51.9 | 57 | 39 | 0.20<br>(239%) |
| KB5 | 0.35987 | 0.5398 | 0.21592 | 97.1 | 75.9 | $1.3\cdot10^{-6}$ | coincide with scalar *ssf* | | | | |
| | 0.19 | 1.31417 | 0.166 | 99.4 | 71 | $3.5\cdot10^{-6}$ | 98.8<br>100.8 | 108.6<br>110.1 | 99 | 108.7 | $3.9\cdot10^{-6}$<br>(11%) |
| $KNbO_3$ | 0.6594 | 1.3188 | 0.4396 | 89.9 | 54.8 | 14.7 | 83.9<br>102.8 | 90<br>90 | 90 | 90 | 15.9<br>(8%) |
| | 3.5303 | 5.2955 | 2.1182 | 70.8 | 252.4 | 18.0 | 83.3<br>100.6 | 90<br>90 | 90 | 90 | 20.4<br>(13%) |
| | 1.0642 | 1.3188 | 0.5889 | 62.9 | 0 | 10.9 | 77.3<br>106.0 | 90<br>90 | 90 | 90 | 17.0<br>(56%) |
| LBO | 0.5321 | 1.0642 | 0.35473 | 90.1 | 142.8 | 0.11 | coincide with scalar *ssf* | | | | |
| | 0.26605 | 1.3188 | 0.22139 | 90.2 | 70.1 | 0.017 | coincide with scalar *ssf* | | | | |
| CBO | 0.5321 | 1.0642 | 0.35473 | 35.3 | 60.0 | 0.41 | 52.6<br>39.0 | 90<br>90 | 43.5 | 90 | 0.58<br>(41%) |
| | 0.35473 | 1.0642 | 0.26605 | 127.7 | 90 | 0.53 | almost coincide with scalar *ssf** | | | | |
| LRB4 | 1.0642 | 1.9079 | 0.6831 | 36.2 | 144.8 | 0.041 | 52.9<br>38.3 | 90<br>90 | 47.7 | 90 | 0.058<br>(41%) |
| | 0.6594 | 1.3188 | 0.4396 | 44.2 | 37.9 | 0.038 | 52.5<br>38.1 | 90<br>90 | 47.7 | 90 | 0.057<br>(50%) |

**Table 3**. Continued

| Crystal | $\lambda_1$, μm (slow) | $\lambda_2$, μm (fast) | $\lambda_3$, μm | Scalar PM | | | Vector PM | | | | |
|---|---|---|---|---|---|---|---|---|---|---|---|
| | | | | Angles, deg. | | $\eta_{scal}^{extr}$, pm²/V² | Angles (pump beams), deg. | | Angles (SFG beam), deg. | | $\eta_{vect}^{extr}$, pm²/V² |
| | | | | θ | φ | | $\theta_p$ | $\varphi_p$ | θ | φ | |
| **type II phase matching (*sff*)** | | | | | | | | | | | |
| KTP | 1.5918 | 1.0642 | 0.6378 | 74.75 | 0 | 2.25 | 90.0<br>83.7 | 0<br>0 | 86.25 | 0 | 2.44<br>(8%) |
| | 1.0642 | 1.5918 | 0.6378 | 53.2 | 0 | 1.47 | 89.9<br>72.0 | 0<br>0 | 83.0 | 0 | 2.44<br>(66%) |
| | 0.76 | 1.77 | 0.5321 | 53.3 | 0 | 1.44 | 90.1<br>68.7 | 0<br>0 | 84.0 | 0 | 2.39<br>(66%) |
| | 2.40688 | 1.90768 | 1.0642 | 54.7 | 0 | 1.60 | 89.9<br>76.0 | 0<br>0 | 82.25 | 0 | 2.53<br>(58%) |
| | 1.90768 | 2.40688 | 1.0642 | 132.7 | 0 | 1.27 | 90.0<br>71.4 | 0<br>0 | 82.0 | 0 | 2.52<br>(98%) |
| KTA | 0.6594 | 1.3188 | 0.4396 | 90.1 | 60.1 | 1.40 | coincide with scalar *sff* | | | | |
| | 1.0642 | 1.9079 | 0.6831 | 49.2 | 0 | 1.74 | 90.1<br>70.2 | 0<br>0 | 83.25 | 0 | 3.24<br>(86%) |
| $KNbO_3$ | 1.0642 | 1.3188 | 0.5889 | 99.0 | 79.8 | $1\cdot10^{-6}$ | 103.7<br>107.3 | 95.4<br>87.4 | 105.2 | 92 | 0.023<br>($2.3\cdot10^6$%) |
| LBO | 0.5321 | 1.0642 | 0.35473 | 42.3 | 90 | 0.062 | coincide with scalar *sff* | | | | |
| CBO | 0.5321 | 1.0642 | 0.35473 | 90.1 | 40.3 | 0.57 | almost coincide with scalar *sff* | | | | |
| | 0.35473 | 1.0642 | 0.26605 | 90.1 | 75.1 | 0.13 | 90<br>90 | 75.6<br>74.2 | 90 | 75.25 | 0.14<br>(8%) |
| LRB4 | 0.6594 | 1.3188 | 0.4396 | 89.5 | 61.3 | 0.039 | almost coincide with scalar *sff* | | | | |
| | 1.0642 | 1.9079 | 0.6831 | 89.5 | 53.5 | 0.053 | almost coincide with scalar *sff* | | | | |

As it is seen from the comparison of data given in Tables 2 and 3, the maximal achievable values of the efficiency η for SHG and SFG usually have the same order of magnitude. The difference is where the wavelengths for SFG and SHG are essentially different, particularly, for *ssf* PM in KB5 and LBO (the second case). For SFG and *sff* PM, using the pump beams of different wavelengths allows achieving PM for the crystals where it is absent for SHG, particularly, for KTA and $KNbO_3$. In a significant number of cases, vector PM ensures higher efficiency of SFG than the scalar one; particularly, for KTP (all cases except the first one of *sff* PM), KTA (except the first case of *sff* PM), $KNbO_3$ (except for the first two cases of *ssf* PM), LBO (the first case of *ssf* PM), CBO (the first case of *ssf* PM) and LRB4 (*ssf* PM) the increase of efficiency is significant and is higher than 1.4 times. It should be noted that although for the $KNbO_3$ crystal the efficiency increase for *sff* PM is very large (23000 times more), the maximal achievable value of η is not significant and is equal to 0.023 pm²/V².

Also note that for KTP crystal in some cases *sff* PM condition can be fulfilled with mutual change of input waves polarizations (*fsf* PM or type III PM [10, 20]). The extreme surfaces for these cases are remarkably different (see, e.g., the surfaces for *sff* PM in KTP, Fig. 5), however, as well as for KTA, evolution of the surface in the transition from one type of phase matching to another can be traced.

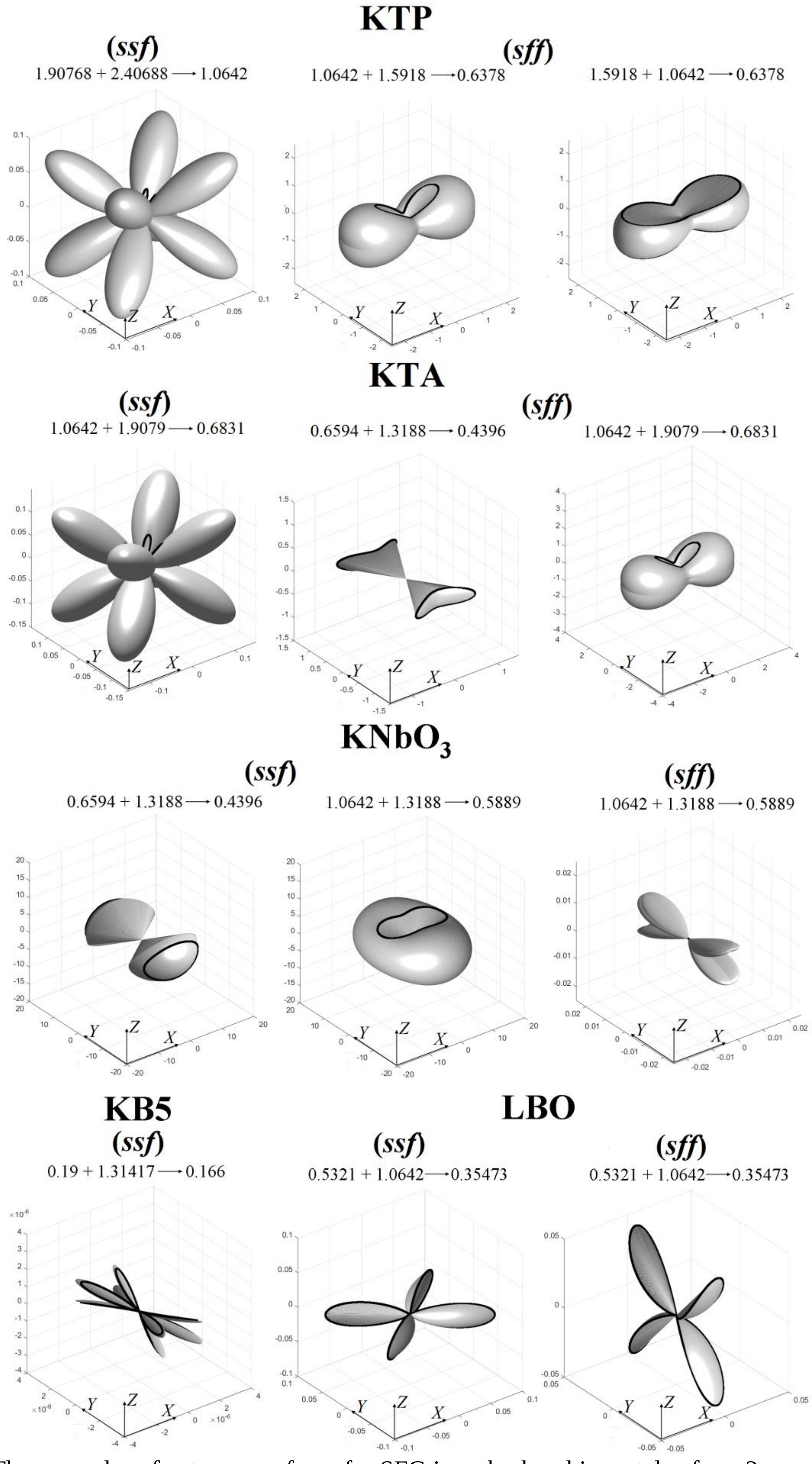

Fig. 5. The examples of extreme surfaces for SFG in orthorhombic crytals of mm2 symmetry (in $pm^2/V^2$).

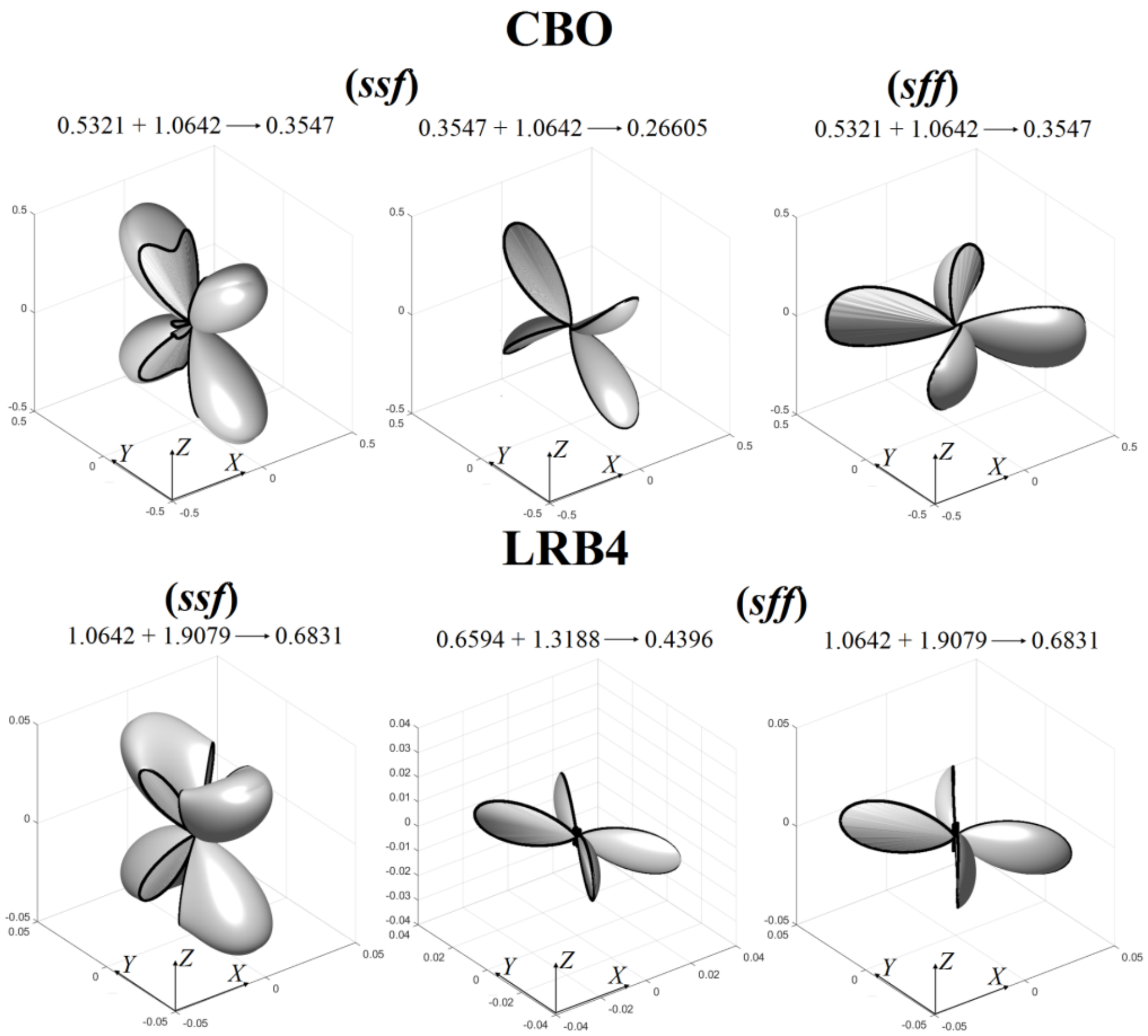

Fig. 6. The examples of extreme surfaces for SFG in orthorhombic crytals of 222 symmetry (in pm$^2$/V$^2$).

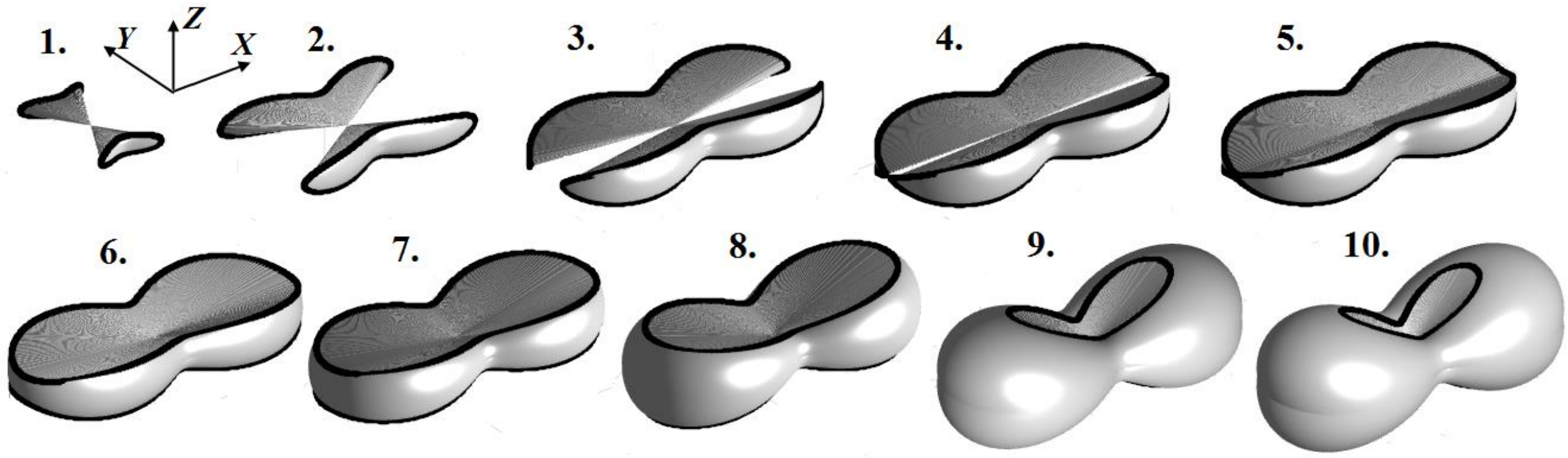

Fig. 7. The evolution of the extreme surface for *sff* PM in KTA crystal (in pm$^2$/V$^2$); 1: (0.6594; 1.3188; 0.4396), 2: (0.6847; 1.3483; 0.4512), 3: (0.6999; 1.3777; 0.4641), 4: (0.7008; 1.3790; 0.46466), 5: (0.7009; 1.3792; 0.4647), 6: (0.7040; 1.3836; 0.4666); (0.710; 1.3924; 0.4702), 7: (0.7404; 1.4366; 0.4886), 9: (0.9023; 1.6723; 0.5861), 10: (1.0642; 1.9079; 0.6831). The scale for all surafaces is the same as for case 10.

*Difference frequency generation*

The results of calculations of the efficiency η for difference frequency generation are given in Table 4, and the examples of extreme surfaces are shown in Fig. 8.

If the Kleiman symmetry rule is strictly followed for the crystals, the maximum achievable efficiency $\eta^{extr}$ for DFG and SFG with the participation of the same sets of wavelengths will be equal. Indeed, according to Kleiman rule, the nonlinear susceptibility tensor is symmetric with respect to index permutations, and therefore the efficiency determined by formula (1) will be the same for the processes $\omega_1 + \omega_2 = \omega_3$, $\omega_3 - \omega_2 = \omega_1$, $\omega_3 - \omega_1 = \omega_2$. Among the considered crystals, this situation occurs for KB5, LBO, CBO and LRB4. For other crystals, the difference in $\eta^{extr}$ is also relatively small,

The highest efficiency increase (tens – thousends of percents) caused by the use of vector PM takes place for KTP (both types of PM), $KNbO_3$, KTA, CBO, LRB4 (type II PM) crystals that agrees with the results obtained for the case of SFG (note that type I PM (*ssf*) in the case of SFG corresponds to type II PM (*fss*) for DFG with the involvement of the same wavelengths etc.).

As it is followed from the comparison of Fig. 8 and Figs. 5,6, some extreme surfaces for DFG are visually more complex than the ones for SFG. Particularly, this difference is observed for the extreme surfaces for *ssf* (SFG) and *fss* (DFG) PM in KTP and KTA. It is similar to the peculiarity observed for $LiIO_3$ crystal in [1] and cause by the fact that the wave vectors of the input beams $\vec{k}_1$ and $\vec{k}_2$ in the case of SFG deviate in opposite directions from the direction of the output beam wave vector $\vec{k}_3$. So if one of these waves is considered as output (DFG), the $\eta_{max}(\theta,\varphi)$ dependency takes more complex form where the maxima of $\eta_{max}$ (in general, non-equivalent) are observed on the both sides of the maximum for SFG as it is shown in Fig. 9 for the case of KTP.

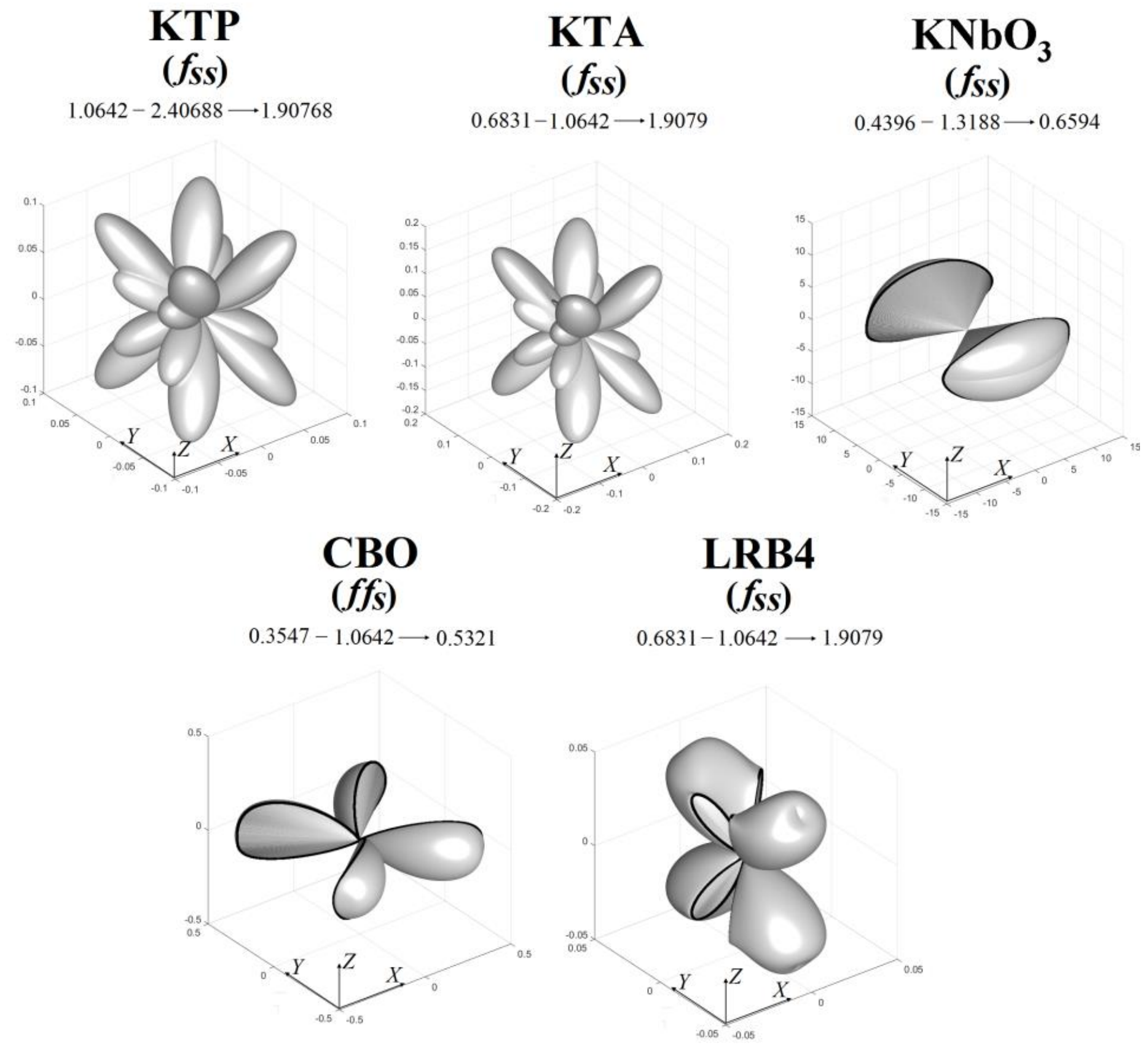

Fig. 8. The examples of extreme surfaces for DFG in nonlinear optical biaxial crytals (in $pm^2/V^2$).

**Table 4** The maximal achievable DFG efficiencies η and corresponding angular parameters

| **Crystal** | **$\lambda_1$, μm** | **$\lambda_2$, μm** | **$\lambda_3$, μm** | **Scalar PM** | | | **Vector PM** | | | | |
|---|---|---|---|---|---|---|---|---|---|---|---|
| | | | | **Angles, deg.** | | $\eta^{extr}_{scal}$, **pm²/V²** | **Angles (pump beams), deg.** | | **Angles (SFG beam), deg.** | | $\eta^{extr}_{vect}$, **pm²/V²** |
| | | | | **θ** | **φ** | | $\theta_p$ | $\varphi_p$ | **θ** | **φ** | |
| **type I phase matching (*ffs*)** | | | | | | | | | | | |
| KTP | 0. 6378 | 1.5918 | 1.0642 | 53.2 | 0 | 1.47 | 96.9<br>107.9 | 0<br>0 | 90 | 0 | 2.44<br>(66 %) |
| | 0.5321 | 1.77 | 0.76 | 53.3 | 0 | 1.44 | 96.1<br>111.4 | 0<br>0 | 90 | 0 | 2.39<br>(66%) |
| | 1.0642 | 2.40688 | 1.90768 | 132.7 | 0 | 1.27 | 98.0<br>108.6 | 0<br>0 | 90 | 0 | 2.52<br>(98.4%) |
| KTA | 0.4396 | 1.3188 | 0.6594 | 90.1 | 60.1 | 1.88 | almost coincide with scalar *ffs* | | | | |
| | 0.6831 | 1.0642 | 1.9079 | 90.0 | 166.0 | 4.14 | coincide with scalar *ffs* | | | | |
| $KNbO_3$ | 0.5889 | 1.3188 | 1.0642 | 103.9 | 58.0 | 0.00095 | 76.1<br>73.2 | 103.6<br>99.8 | 78.3 | 106.5 | 0.028<br>(2847%) |
| | 1.9105 | 4.7762 | 3.1841 | 79.9 | 125.4 | $5 \cdot 10^{-5}$ | 76.8<br>74.5 | 97.1<br>92.7 | 78.3 | 99.8 | 0.021<br>(41900%) |
| **type II phase matching (*fsf,fss*)** | | | | | | | | | | | |
| KTP | 0.6378 | 1.5918(s) | 1.064(f) | 74.8 | 0 | 2.25 | 93.7<br>89.9 | 0<br>0 | 96.3 | 0 | 2.44<br>(8%) |
| | 0.6378 | 1.5918(s) | 1.064(s) | 37.7 | 33.3 | 0.033 | 57.1<br>48.0 | 140.9<br>129.2 | 63.5 | 147.3 | 0.12<br>(263.6%) |
| | 0.5321 | 1.77(s) | 0.76(s) | 139.0 | 144.9 | 0.034 | 119.5<br>130.7 | 141.6<br>126.8 | 114.5 | 146.8 | 0.15<br>(341%) |
| | 1.0642 | 2.40688(s) | 1.90768(f) | 54.7 | 0 | 1.60 | 82.3<br>89.9 | 0<br>0 | 76 | 0 | 2.53<br>(58%) |
| | 1.0642 | 2.40688(s) | 1.90768(s) | 147.1 | 30.4 | 0.025 | 55.0<br>46.2 | 39.4<br>50.7 | 62.5 | 32.3 | 0.12<br>(380%) |
| KTA | 0.4396 | 1.3188(s) | 0.6594(s) | 121.0 | 135.6 | 0.070 | coincide with scalar *fss* | | | | |
| | 0.6831 | 1.0642(s) | 1.9079(f) | 49.2 | 0 | 1.74 | 83.3<br>90.1 | 0<br>0 | 70.3 | 0 | 3.24<br>(86%) |
| | 0.6831 | 1.0642(s) | 1.9079(s) | 36.6 | 36.7 | 0.078 | 56.2<br>61.8 | 140.4<br>146.1 | 46.8 | 127.8 | 0.22<br>(182%) |
| $KNbO_3$ | 0.4396 | 1.3188(s) | 0.6594(s) | 90.1 | 141.5 | 13.7 | almost coincide with scalar *fss* | | | | |
| | 2.1182 | 3.5303(s) | 5.2955(s) | 86.0 | 0 | 17.0 | 90.1<br>96.8 | 90<br>90 | 79.5 | 90 | 17.8<br>(4.7%) |
| | 0.5889 | 1.064(s) | 1.3188(f) | 110.2 | 71.8 | 0.00012 | 75.5<br>77.4 | 96<br>99.6 | 73 | 92 | 0.022<br>(18233%) |
| | 0.5889 | 1.064(s) | 1.3188(s) | 62.7 | 0 | 11.8 | 90<br>102.8 | 90<br>90 | 74 | 90 | 14.8<br>(25.4%) |
| | 1.9105 | 4.7762(s) | 3.1841(s) | 66.9 | 0 | 14.4 | 89.9<br>104.3 | 90<br>90 | 80.8 | 90 | 17.3<br>(20%) |

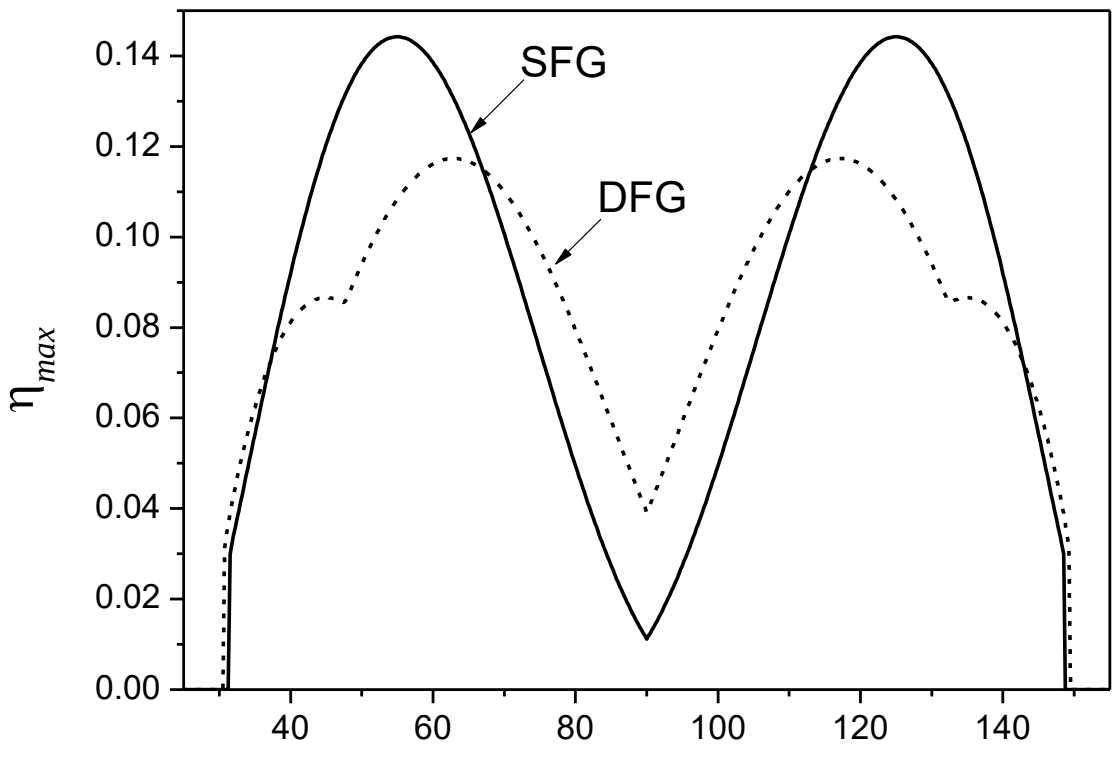

Fig. 9. The cross-section of the extreme surfaces for KTP crystal at φ = 39.2° that corresponds to the maximum of the efficiency for *ssf* PM (SFG). The wavelengths of the 'slow' waves are 1.90768 and 2.40688 μm, the wavelength of the 'fast' wave is 1.0642 μm.

## Conclusions

The extreme surfaces technique is used for the determination of the maximal achievable efficiency η of second harmonic, sum and difference frequency generation in several biaxial orthorhombic nonlinear optical crystals, i.e., KTP, KTA, KB5, $KNbO_3$, LBO, CBO, LRB4. Both the cases of vector and scalar phase matching are analyzed and compared for all crystals. The optimal geometries of vector phase matching, i.e., the directions of wave vectors of the pump and output beams ensuring the highest possible values of the efficiency are determined. As it is shown, in a significant number of cases, vector phase matching ensures higher efficiencies than the scalar one. Particularly, it takes place for second harmonic generation (type I or *ssf* phase matching) in KB5 (the relative increase of the efficiency is 145%), KTP (49%), $KNbO_3$ (39%) and KTA (17%). For sum frequency generation the different sets of wavelengths were used for calculation of the maximal efficiency for considered crystals and, as it is shown, the maximal values of η strongly depends on them. In general, the highest increase of the efficiency (higher than 1.4 times) in comparison with the scalar case is observed for KTP, KTA, $KNbO_3$, LBO (*ssf* PM), CBO (*ssf* PM) and LRB4 (*ssf* PM). Remarkably, the extremal increase in efficiency (around 39000 times) occurs for $KNbO_3$; however, it is mainly caused by low values of scalar phase matching efficiency in this crystal. The situation for difference frequency generation is, in general, similar to the one for the case of sum frequency generation. The obtained results can be used for enhancement of the performance of nonlinear optical devices.

## Funding

This research has received funding from the European Union's Horizon 2020 research and innovation programme under the Marie Skłodowska-Curie grant agreement No 778156 and from Ministry of Education and Science of Ukraine in the frames of projects 'Nanomatrix' (0122U000951) and 'Nanoelectronics' (0123U101695).